\documentclass{elsart}

\usepackage{amssymb}

\begin{document}

\begin{frontmatter}



\title{Hidden symmetries in two dimensional field theory}


\author{Michael Creutz}

\address{Physics Department, Brookhaven National Laboratory\\
Upton, NY 11973, USA}

\begin{abstract}
The bosonization process elegantly shows the equivalence of massless
scalar and fermion fields in two space-time dimensions.  However, with
multiple fermions the technique often obscures global symmetries.
Witten's non-Abelian bosonization makes these symmetries explicit, but
at the expense of a somewhat complicated bosonic action.  Frenkel and
Kac have presented an intricate mathematical formalism relating the
various approaches.  Here I reduce these arguments to the simplest
case of a single massless scalar field.  In particular, using only
elementary quantum field theory concepts, I expose a hidden
$SU(2)\times SU(2)$ chiral symmetry in this trivial theory.  I then
discuss in what sense this field should be interpreted as a Goldstone
boson.
\end{abstract}

\end{frontmatter}


\section{Introduction}

A large variety of two dimensional models can be related and often
solved via the process of bosonization.  \cite
{Skyrme:1961vr,Coleman:1975pw,Mandelstam:1975hb,Halpern:1975nm,Ha:1984ph}
This process, however, often obscures certain symmetries. For example,
the two flavor generalization \cite{Coleman:1976uz,Halpern:1975jc} of
the Schwinger model \cite{Schwinger:1962tp} in the fermion formulation
has an $SU(2)\times SU(2)$ chiral symmetry, but the bosonic solution
has one massive and one massless scalar field, both free.  In the
strong coupling limit the massive particle should become irrelevant,
leaving the puzzle of how can a chiral symmetry appear in the trivial
theory of only a single free massless scalar.

This question has been discussed in terms of another formulation of
bosonization, where the basic fields are elements of a group, and the
chiral symmetries involve rotations of these elements.  This
``non-Abelian'' bosonization \cite{Witten:1983ar} involves a chiral
Lagrangian containing a rather interesting topological term.
\cite{Wess:1971yu,Witten:1983tw} While this formulation keeps the
chiral symmetries more transparent, the mapping between the chiral
fields and the alternative Abelian bosonization is somewhat obscure.
Some of the connections were discussed in a series of papers by
Affleck, \cite{Affleck:1985wb,Affleck:1985wa} and an explicit
construction of the connection is given by Amaral and
StephanyRuiz.\cite{doAmaral:1991em} Halpern \cite{Halpern:1975jc}
gives the form for the chiral currents in the multi-flavor Schwinger
model.

In this paper I return to this old topic with further discussion of
how a non-Abelian current algebra is hidden in the simplest two
dimensional scalar field theory.  The construction is well known to
the string theory community and is a special case of the general
technique of Frenkel and Kac.
\cite{Frenkel:1980rn,Goddard:1988fw,Halpern:1995js} That discussion,
however, is placed in a rather formal context; my goal here is to
elucidate the surprising consequences more transparently, specializing
to the simplest case and using only concepts from elementary quantum
field theory.

After a discussion of the connection with Witten's non-Abelian
formulation, I turn to some comments on the role and counting of
Goldstone bosons.  In two dimensions the definition of a Goldstone
boson is subject to some interpretation.  On the one hand, infrared
fluctuations preclude the matrix valued fields from acquiring a vacuum
expectation value.  This is the basis of the familiar arguments that
spontaneous breaking of a continuous symmetry cannot occur in two
dimensions.\cite{Coleman:1973ci} On the other hand, the chiral charges
are rather singular objects, and with a simple cutoff the vacuum is
not annihilated by them, even in the limit that the cutoff is removed.
The latter is sufficient to require the existence of a massless
particle in the spectrum, and forms the basis of one proof of the
Goldstone theorem.\cite{goldstone} In this sense a two dimensional
field theory can exhibit a Goldstone boson, although it must be free.

Motivated by the simplest case of the strongly coupled two flavor
Schwinger model, I will show that the trivial field theory of a free
massless boson in one space dimension has a hidden $SU(2)\times SU(2)$
symmetry.  In particular, I will construct conserved currents
$J_{R,\mu}^\alpha(x)$ and $J_{L,\mu}^\alpha(x)$, where $\alpha$ is an
``isospin'' index running from 1 to 3, $\mu\in\{0,1\}$ is a Lorentz
index, and $L,R$ label left and right handed parts.  The resulting
charge densities satisfy the equal time commutation relations
\begin{eqnarray}
\label{commutator}
[J_{R,0}^\alpha(x),J_{R,0}^\beta(y)]&
=&i\epsilon^{\alpha\beta\gamma}J_{R,0}^\gamma(x)\ \delta(x-y)
+A \delta^{\alpha\beta}\partial_x \delta(x-y),\cr
[J_{L,0}^\alpha(x),J_{L,0}^\beta(y)]&
=&i\epsilon^{\alpha\beta\gamma}J_{L,0}^\gamma(x)\ \delta(x-y)
-A \delta^{\alpha\beta}\partial_x \delta(x-y),\cr
[J_{R,0}^\alpha(x),J_{L,0}^\beta(y)]&=&0.
\end{eqnarray}
The notation left or right indicates that the corresponding currents
are to be constructed from operators involving particles moving to the
right or left, respectively.  I include here a Schwinger
\cite{Schwinger:1959xd} term with coefficient $A$ that will be
determined.  The Schwinger terms for the two chiralities differ in
sign, assuring that they cancel in the commutators of the vector
current.

I organize this paper as follows.  In Section \ref{introduction} I
make a few remarks on why these two dimensional models may be useful
for the understanding of chiral symmetry in four dimensions.  Section
\ref{scalar} I establish notation by defining the Hilbert space in
which I work.  Section \ref{chiral} discusses splitting the massless
fields into left and right moving parts.  With Section
\ref{exponentiation} I review the concept of normal-ordered
exponentiated fields.  The basic result for the currents appears in
Section \ref{currents}.  I then turn to the connection with group
valued fields and the corresponding equations of motion in Section
\ref{group}.  Section \ref{goldstone} explores the question of whether
the underlying massless field can be thought of as a Goldstone boson.
A few final comments appear in Section \ref{conclusions}.

\section{Four dimensional motivations}
\label{introduction}

Of course we live in a four dimensional world; so, it is perhaps worth
mentioning some of the reasons these two dimensional models are worth
studying.  Two dimensional field theories are of intense interest to
the string theory community, where the essense of these results is
well known.  However, my interest comes from a rather different
direction, related to attempts to formulate quantum field theory
beyond the realm of perturbation theory.

For the strong interactions, spontaneous breaking of a global chiral
symmetry plays a major role in the understanding of the hadronic
spectrum.  Indeed, the lightness of the pions relative to the rho
mesons has long been explained in this framework.  However, the
phenomenon is inherently not perturbative.  The lattice provides the
primary non-perturbative method in field theory, but issues related to
anomaly cancellation make that approach rather complicated.
\cite{Creutz:2000bs} Effective chiral Lagrangians provide another
powerful route to non-perturbative information, although in a less
quantitative manner due to increasing numbers of arbitrary parameters
at higher order.  To understand these issues better, the solvable
models in two dimensions can provide insight into how chiral symmetry
works.  The symmetries discussed here are also in the four dimensional
quark-gluon theory with two massless flavors, although the counting of
Goldstone bosons manifests itself somewhat differently.  While much of
the rigorous mathematical work on these two dimensional models is
built on conformal symmetry, this is of less relevance in four
dimensions.  Indeed, one of the remarkable properties of quark
confining dynamics is how asymptotic freedom manages to avoid the
conformal symmetry of the classical theory with massless quarks.

Understanding chiral symmetry is presumably an important step towards
a non-per\-tur\-bative formulation of chiral gauge theories.  As the
weak interactions do not conserve parity, we know that the gauge
fields are coupled in a chirally non-symmetric way to the fundamental
fermions.  Without a non-perturbative formulation, one might worry
whether the standard model is well defined as a field theory.  But
here the lattice issues are unresolved; a fully finite lattice
regularization that preserves an exact underlying gauge symmetry
remains elusive.  This is in contrast to the bosonization technique,
around which this discussion revolves, where there are no problems
defining generalizations of the Schwinger model to a chiral theory, as
long as anomalies are properly cancelled.\cite{Halliday:1985tg}
Indeed, this shows that the absence of a clean lattice regulator does
not preclude the existence of at least these simplified chiral
theories.

\section{The scalar field}
\label{scalar}

To establish notation, in this section I set up the basic scalar field
theory in terms of which I will construct the non-Abelian currents.  I
work at a fixed time in a Hilbert space formulation.  The states of
this space are generated by bosonic creation operators $a^\dagger_p$
operating on a normalized vacuum state $|0\rangle$.  The momentum
space commutation relations are the usual
\begin{equation}
[a_p,a^\dagger_{p^\prime}]=4\pi p_0\ \delta(p-p^\prime),
\end{equation}
and the vacuum is annihilated by the destruction operators,
$a_p|0\rangle=0$.  As I have in mind the massless theory, I take
$p_0=|p|$.  The local field and its conjugate momentum are 
\begin{eqnarray}
\Phi(x)=\int_{-\infty}^\infty {dp\over 4\pi p_0}\left( e^{-ipx} a_p +
e^{ipx} a^\dagger_p\right),\cr
\cr
\Pi(x)=i\int_{-\infty}^\infty {dp\over 4\pi}\left( e^{-ipx} a_p -
e^{ipx} a^\dagger_p\right).
\end{eqnarray}
These satisfy the canonical position space commutation relations
$$
[\Pi(x),\Phi(y)]=i\delta(x-y).
$$ For a massless particle, time evolution is given by the simple free
field Hamiltonian
\begin{equation}
H=\int dx\ (:\Pi^2(x):+:(\partial_x\Phi(x))^2:)
=\int {dp\over 4\pi} a^\dagger_p a_p.
\label{hamiltonian}
\end{equation}
The colons denote normal ordering with respect to the creation and
annihilation operators, i.e. all annihilation operators are placed to
the right of all creation operators.  This normal ordering ensures a
zero energy vacuum.

As is well known,\cite{Coleman:1973ci} a massless field in two
dimensions is a rather singular object.  In particular, the two point
function $\langle 0| \Phi(x)\Phi(y) |0\rangle$ has an infrared
divergence.  This can be circumvented by considering correlations
between derivatives of the field, which are better behaved.  I will
shortly introduce infrared and ultraviolet cutoffs giving well defined
field correlators.  Any final conclusions require combinations of the
fields having a a finite limit as these cutoff parameters are removed.

\section{Chiral fields}
\label{chiral}

In one dimension there is a natural notion of chirality for massless
particles.  A particle going to the right in one frame does so at the
speed of light in all frames.  This is true regardless of whether the
particle is a boson or a fermion.  Thus it is natural to separate the
field into right and left moving parts
\begin{equation}
\Phi(x)=\Phi_R(x)+\Phi_L(x),
\end{equation}
where the right handed piece only involves operators for positive momentum
\begin{equation}
\Phi_R(x)=\int_0^\infty {dp\over 4\pi p_0}\left( e^{-ipx} a_p +
e^{ipx} a^\dagger_p\right).
\end{equation}
Correspondingly, the left handed field only involves negative
momentum.  The goal here is to construct the right (left) handed
currents using only the right (left) handed field.

Note that the canonical momentum satisfies
\begin{equation}
\Pi(x)=-\partial_x(\Phi_R(x)-\Phi_L(x)).
\end{equation}
Thus one can alternatively work with $\{\Pi(x),\Phi(x)\}$ or
$\{\Phi_R(x),\Phi_L(x)\}$ as a complete set of operators in the
Hilbert space.  Also note that formally $\Phi_R(x)$ does not commute
with itself at different positions.  However derivatives of the field
do, and, as mentioned above, only derivatives of the field are
physically sensible.  With this proviso, either the left or right
fields define a relativistic quantum field theory on their own.  Were
a mass present, the left and right fields would mix under Lorentz
transformations and should not be considered independently.

Under the Hamiltonian of Eq.~(\ref{hamiltonian}), the equations of
motion for the chiral fields are particularly simple.  The right
(left) field only creates right (left) moving waves, or in equations
\begin{eqnarray}
(\partial_t+\partial_x)\Phi_R(x)=0,\cr
(\partial_t-\partial_x)\Phi_L(x)=0.
\end{eqnarray}
For the time being I will concentrate on the right handed field.

As with the full field, correlation functions of these fields are
divergent.  To get things under better control, I introduce an
infrared cutoff $m$ and an ultraviolet cutoff $\epsilon$ with the
definition
\begin{equation}
\Phi_R(x)=\int_m^\infty {dp\ {e^{-\epsilon p/2}}\over 4\pi
p}\left( e^{-ipx} a_p + e^{ipx} a^\dagger_p\right).
\end{equation}
Both cutoffs are to be taken to zero at the end of any calculation
of physical relevance.

There is some arbitrariness in both these cutoffs.  In particular,
another popular infrared cutoff gives the scalar boson a small
physical mass via the choice $p_0=\sqrt{p^2+m^2}$.  All the following
discussion could be done either way.  A physical mass, however,
complicates the separation of chiral parts, since Lorentz
transformations will mix them.  With the choice taken here, the left
and right movers remain independent, although Lorentz transformations
will change the cutoff.

With the cutoffs in place, the correlation of two of these operators
becomes well defined
\begin{eqnarray}
\Delta_R(x-y)=\langle 0 | \Phi_R(x) \Phi_R(y)|0\rangle
=\int_m^\infty {dp\ {e^{-\epsilon p}}\over 4\pi
p} e^{-ip(x-y)}\cr
={1\over 4\pi}\left(C-\log(x-y-i\epsilon)-\log(m)-{i\pi\over 2}\right).
\label{propagator}
\end{eqnarray}
Here $C$ is the Gompertz constant\cite{gompertz} divided by $e$ and
has the value
\begin{equation}
C=\int_1^\infty {dp\ {e^{-p}}\over p}=0.21938\ldots.
\end{equation}

Note that this ``propagator'' diverges logarithmically as $m$ goes to
zero, although its derivatives do not.  For example,
\begin{equation}
\langle 0| \partial_x \Phi_R(x) \partial_y \Phi_R(y) |0\rangle
={-1\over 4\pi (x-y-i\epsilon)^2}
\label{deriv}
\end{equation}
remains a tempered distribution as $\epsilon$ goes to zero.


\section{Exponentiated fields}
\label{exponentiation}

The usual construction of fermionic operators in the bosonization
process involves exponentiated scalar fields.  This will also be the
case for the currents below, although detailed factors will differ.
To keep things well defined, I consider the normal-ordered operator
with the cutoffs in place
\begin{equation}
:e^{i\beta\Phi_R(x)}:
\ =\ \exp\left(i\beta\int_m^\infty 
{dpe^{-\epsilon p/2}\over 4\pi p} e^{ipx}a^\dagger_p\right) 
\exp\left(i\beta\int_m^\infty 
{dpe^{-\epsilon p/2}\over 4\pi p} e^{-ipx}a_p\right). 
\end{equation}
The expression in Eq.~(\ref{propagator}) for the propagator gives a
rather simple relation to normal order the product of two of these
operators
\begin{eqnarray}
:e^{i\beta\Phi_R(x)}:\ :e^{i\beta^\prime\Phi_R(y)}:
\ =\ :e^{i\beta\Phi_R(x)+i\beta^\prime\Phi_R(y)}:
\ \exp(-\beta\beta^\prime \Delta_R(x-y))\cr
\cr
=:e^{i\beta\Phi_R(x)}e^{i\beta^\prime\Phi_R(y)}:
\left(
{-ie^{C}\over m(x-y-i\epsilon)}
\right)^{-\beta\beta^\prime/4\pi}.
\label{product}
\end{eqnarray}
I will always be working with $\beta\beta^\prime$ an integer multiple
of $4\pi$; thus, there is no phase ambiguity.  This will be the key
relation in the following.

In two dimensions the free massless fermion propagator is proportional
to $1/(x-y)$.  This is the basis of the usual bosonization which takes
$\beta=2\sqrt \pi$ and identifies
\begin{equation}
\psi_R(x)= {e^{C/2}\over \sqrt{2\pi} }
\ \lim_{m\rightarrow 0}\sqrt m :e^{2i\sqrt\pi \Phi_R(x)}:.
\end{equation}
The factor in front gives conventionally normalized fermionic
commutation relations, and an additional phase appears between the
left and right handed fermion fields to have them anticommute.
However my goal here is not the fermion field, but rather the
non-Abelian currents.

Note that if $-\beta\beta^\prime=8\pi$ the spatial dependence in
Eq.~(\ref{product}) is proportional to that of the correlator between
two derivatives of the field as given in Eq.~(\ref{deriv}).  This lies
at the heart of the construction of the currents in the next section.

\section{The currents}
\label{currents}

I concentrate here on constructing the right handed current from the
right handed field.  The construction of the left handed current
proceeds in a parallel fashion.  I start by selecting one component of
the isovector current as the trivially conserved
\begin{equation}
J^3_{R,\mu}=-k\epsilon_{\mu\nu}\partial_\nu \Phi_R(x).
\end{equation}
Here $\epsilon_{\mu\nu}$ is the antisymmetric tensor with
$\epsilon_{0,1}=1$ and $k$ is a normalization factor that will shortly
be determined.  Note that because this involves derivatives,
correlation functions of this current are well defined tempered
distributions.  My convention on repeated Lorentz indices is
understood as a summation with the metric $g_{00}=-g_{11}=1$.  Thus
the charge density is $J^3_{R,0}(x)=k\partial_x \Phi_R(x)$, and the
associated charge is
\begin{equation}
Q^3=\int_{-\infty}^\infty dx\ J^3_{R,0}(x)= k(\Phi_R(\infty)-\Phi_R(-\infty)).
\label{charge}
\end{equation}
The commutator of the current with itself gives the coefficient of the
Schwinger term
\begin{eqnarray}
[J^3_{R,0}(x),J_{R,0}^3(y)]=k^2\partial_x\partial_y
\left(\Delta(x-y)-\Delta(y-x)\right)\cr 
\cr
={k^2\over 4\pi}\partial_x
\left( {1\over x-y-i\epsilon}-{1\over x-y+i\epsilon}\right)
\rightarrow {k^2\over 2}\partial_x\delta(x-y).
\label{schwinger}
\end{eqnarray}
This relates the coefficient in Eq.~(\ref{commutator}) to the
normalization k (still to be determined), $A={k^2\over 2}$.

For the other components of the currents it is easiest to work with
raising and lowering combinations
$J^\pm_{R,\mu}={1\over\sqrt2}(J^1_{R,\mu}\pm iJ^2_{R,\mu})$.  Then the
desired commutation relations reduce to
\begin{eqnarray}
[J^3_{R,0}(x),J^\pm_{R,0}(y)]&=&\pm J^\pm_{R,0}(x)\ \delta(x-y),\cr
[J^+_{R,0}(x),J^-_{R,0}(y)]&=&J^3_{R,0}(x)\ \delta(x-y)
+A\partial_x\delta(x-y).
\label{ladders}
\end{eqnarray}
The first relation in conjunction with Eq.~(\ref{charge}) indicates
that $J^+$ must induce a ``kink'' of size $1/k$ in the field $\Phi_R$.
The fermionic operators in Abelian bosonization do something similar,
but the size of the kink differs.  This observation suggests I try the
form
\begin{equation}
J_{R,0}^+(x)\ \sim\ :e^{i\beta\Phi_R(x)}:.
\end{equation}
Since the currents should commute at non-vanishing separation, I
should take $\beta^2=8\pi n$ with $n$ an integer.  For the remainder
of this discussion I take the lowest value
\begin{equation}
\beta=2\sqrt{2\pi}.
\end{equation}
This is the square root of two times the value taken to construct
fermion fields.  This follows intuitively from the fact that the
current should be a fermion bilinear with two ``kinks'' in orthogonal
directions.

For the first commutator in Eq.~(\ref{ladders}) look at
\begin{eqnarray}
[ \partial_x \Phi_R(x),:e^{i\beta\Phi_R(y)}:]
=i\beta\ :e^{i\beta \Phi_R(y)}:\ &
\partial_x
(\Delta(x-y)-\Delta(y-x))\cr
= { \sqrt{2\pi}}:e^{i\beta\Phi_R(y)}:\ \delta(x-y). &
\end{eqnarray}
This says I should take $k=1/\sqrt{2\pi}$ and
\begin{equation}
J^3_{R,0}(x)={1\over \sqrt{2\pi}} \partial_x\Phi_R(x).
\label{j3}
\end{equation}

To get the normalization of the other currents, use
Eq.~(\ref{product}) to work out the commutator
\begin{eqnarray}
[\ :e^{i\beta\Phi_R(x)}:,\ :e^{-i\beta\Phi_R(y)}:\ ]
\ =&\cr
:e^{i\beta(\Phi_R(x)-\Phi_R(y))}:\ 
{-e^{2C}\over m^2}&
\left(
{1\over (x-y-i\epsilon)^2}-{1\over (x-y+i\epsilon)^2}
\right).
\end{eqnarray}
As $\epsilon$ becomes small the last factor becomes a derivative of a
delta function
\begin{equation}
{1\over (x-y-i\epsilon)^2}-{1\over (x-y+i\epsilon)^2}
\rightarrow {-2\pi i}{d\over dx}\delta(x-y).
\end{equation}
Using the Leibnitz rule
\begin{equation}
f(x)\ \delta^\prime(x)=-f^\prime(0)\ \delta(x)+f(0)\ \delta^\prime(x),
\end{equation}
the commutator reduces to 
\begin{equation}
[ :e^{i\beta\Phi_R(x)}:,:e^{-i\beta\Phi_R(y)}:]
= {2\pi e^{2C}\over m^2} \left( 
4\pi \delta(x-y)J_{R,0}^3(x) 
+i\delta^\prime(x-y)
\right).
\end{equation}
Absorbing the prefactors, the desired commutation follows with
\begin{equation}
J^+_{R,0}(x)={e^{-C} \over 2\pi\sqrt{2}}\ \lim_{m\rightarrow 0}
m :e^{i\beta\Phi_R(x)}:.
\end{equation}
With Eq.~(\ref{j3}), this is the final result.  The Schwinger term
appears with the same coefficient as in Eq.~(\ref{schwinger}), serving
as a consistency check on the normalizations.  Finally, note that the
equations of motion give the spatial component of the currents
$J_{R,1}^\alpha=J_{R,0}^\alpha$ (For the left currents there is a
relative minus sign in this relation).

\section{Matrix fields}
\label{group}

The non-Abelian bosonization of Witten is formulated in terms of a
matrix valued field in the fundamental representation of the symmetry
group.  As I now have the basic symmetry generators, I should be able
to construct this matrix field from the elementary scalar as well.
This section outlines the procedure.  The construction gives rise to
an equation of motion that contains a Wess-Zumino term.

Motivated by Witten's\cite{Witten:1983ar} discussion, I start by
looking for a matrix valued field as a product
\begin{equation}
g=g_Lg_R,
\end{equation}
 where $g_R$ ($g_L$) is constructed from right (left) fields alone.  I
 ask that these satisfy the equations
\begin{eqnarray}
\partial_x g_R=g_R\sigma^\alpha J_{R,0}^\alpha\cr
\partial_x g_L=\sigma^\alpha J_{L,0}^\alpha g_L.
\end{eqnarray}
These are solved as ``$X$ ordered'' integrals,\cite{Curtright:1994be}
{\it i.e.}
\begin{equation}
g_R(x)=X\left(\exp \left(i\int_{-\infty}^x dx^\prime\ \sigma^\alpha
J_R^\alpha(x^\prime)\right)\right).
\end{equation}
The left field would be formulated in the corresponding way with an an
``anti-$X$ ordered'' integral.  This is in direct analogy with path
ordered products of group elements in gauge theory.  I assume here
that the $X$ ordered integration starts with the unit matrix at
$-\infty$.  Later I will argue that the details of the boundary
condition are unimportant since the correlations of these matrix
fields decrease to zero with separation.

Being constructed only from right moving fields, $g_R$ should satisfy
the relation $(\partial_t+\partial_x) g_R=0$.  Similarly the left
field satisfies $(\partial_t-\partial_x) g_L=0$.  For the product we
thus wind up with the equation of motion
\begin{equation}
\partial_\mu\partial_\mu g=((\partial_t+\partial_x)g)g^\dagger
((\partial_t-\partial_x)g
=(\partial_\mu g)g^\dagger\partial_\mu g+\epsilon_{\mu\nu}
(\partial_\mu g)g^\dagger\partial_\nu g.
\end{equation}
The piece involving $\epsilon_{\mu\nu}$ is the Wess-Zumino term.
Witten\cite{Witten:1983ar} has extensively discussed how to obtain
these equations of motion from a Lagrangian including a topological
term.

Note that given the field $g$, it is straightforward to go back and
reconstruct the currents
\begin{eqnarray}
j^\alpha_{R,\mu}={-1\over 4}
{\rm Tr}\sigma^\alpha g^\dagger 
(\epsilon_{\mu\nu}\partial_\nu+\partial_\mu) g\cr
j^\alpha_{L,\mu}={-1\over 4}
{\rm Tr} g^\dagger \sigma^\alpha 
(\epsilon_{\mu\nu}\partial_\nu-\partial_\mu) g.
\end{eqnarray}

\section{Goldstone bosons}
\label{goldstone}

My starting point was a massless field.  Is it a Goldstone boson?
This question is subtle due to the borderline nature of two
dimensions.  Mermin and Wagner\cite{Mermin:1966fe} showed that in one
space dimension one cannot have ferromagnetism in the sense of an
order parameter with a continuous symmetry acquiring an expectation
value.  Coleman\cite{Coleman:1973ci} proved this result in the
framework of relativistic quantum field theory, relating it to the
singular nature of the propagator for a massless field.
Affleck\cite{Affleck:1985wa} rephrases the Mermin Wagner argument in
terms of the correlator of two matrix valued fields decreasing to zero
at long distances. 

On the other hand, the basic proof of the Goldstone
theorem\cite{goldstone} considers a symmetry generator that does not
annihilate the vacuum.  If you have a such a charge that commutes with
the Hamiltonian, there must exist states of arbitrarily low energy.
These can be created by applying the local charge density times a
slowly varying test function on the vacuum.  In a theory of particles,
states of arbitrarily low energy are built from massless particles
moving with arbitrarily low momentum.  These are the Goldstone bosons.

So, in this massless theory, do the relevant charges annihilate the
vacuum?  Generally global charges are somewhat singular objects due to
infrared issues.\cite{orzalesi} Local charge densities, being
derivatives of the field, are well defined.  Working directly with
them I consider damping the third charge with a Gaussian factor
\begin{equation}
Q^3_R=\int dx e^{-\alpha x^2} J_{R,0}^3(x)
={-i\over \sqrt{2\alpha}}
\int_0^\infty {dp\over 4\pi} e^{-p^2/ 4\alpha}(a_p-a_p^\dagger)
\end{equation}
As this is linear in the creation/annihilation operators, its vacuum
expectation value vanishes.  However, its application on the vacuum
does not itself vanish, as can be seen from evaluating
\begin{equation}
\langle 0 | Q^3 Q^3|0\rangle
={1\over 8\pi\alpha}
\int_0^\infty {p\ dp} e^{-p^2/2\alpha}
={1\over 8\pi}.
\end{equation}
The damping factor cancels out, giving a finite non-vanishing result
as it is removed.  To the extent that this limit defines the charge,
it does not annihilate the vacuum.  This is enough to show that states
of arbitrarily low energy must exist.  In particular, the state
$Q_3|0\rangle$, with the cutoff $\alpha$ above in place, is orthogonal
to the vacuum but has an expectation value for the Hamiltonian that
goes to zero as $\alpha$ does, i.e.
\begin{equation}
\langle 0 | Q^3 H Q^3|0\rangle\
={1\over 8\pi\alpha}
\int_0^\infty {p^2\ dp} e^{-p^2/2\alpha}
={\sqrt\alpha\over 8\sqrt{2\pi}}
\ \rightarrow\ 0.
\end{equation}
These states are long wavelength modes of the starting massless field,
and by this interpretation it is indeed a Goldstone boson.

The full current algebra consists of three isospin currents.  However,
this borderline case of two dimensions allows a single boson to suffice
for all.  This contrasts with higher dimensions where spontaneous
symmetry breaking is better defined and there are three independent
Goldstone bosons for the $SU(2)$ case.

The Merman-Wagner and Coleman discussions point out that, unlike in
higher dimensions, this breaking of symmetry does not appear in the
expectation value of the group valued field $g$.  In more dimensions
$g$ is usually written as the exponentiated Goldstone boson field.
But in the two dimensional case the scalar field has infinite
fluctuations, as seen in Eq.~(\ref{propagator}).  These infinite
fluctuations make the expectation value of $g$ vanish.

\section{Final comments}
\label{conclusions}

Note that this discussion pays little attention to boundary
conditions.  Indeed, the fact that the expectation value of $g$
vanishes suggests that they are irrelevant.  This is in some contrast
to the usual discussion of the Wess-Zumino term where working with a
compact space is a basic starting point.

Turning on a small common mass for the fermions in the two flavor
Schwinger model will drive $g$ to have an expectation value.  Then the
low energy spectrum of the theory will indeed have three degenerate
light mesons.  One is from the fundamental field $\Phi$ and represents
the neutral pion in analogy to the four dimensional theory.  The other
two light excitations are solitons representing the charged
pions\cite{Creutz:1994px}.

Given the singular nature of a massless field and the ability to map
between rather different looking formulations, one might ask if the
number of massless particles is a well defined concept in two
dimensions.  A simple physical argument that precisely counts the
bosons in a theory is to calculate the vacuum energy per unit volume
at finite temperature.  This gives a Stefan-Boltzmann law, which in
two dimensions reads
\begin{equation}
<E/V>={n_b\pi T^2\over 6}
\end{equation}
where $n_b$ is the number of massless particles.  Thus the starting
theory has indeed only one Goldstone boson.

Note that this finite temperature energy density can be calculated
either in terms of the boson field or in terms of the equivalent
fermion field.  In the latter case the Fermi-Dirac statistics gives a
factor of two reduction, but that is cancelled by the presence of both
particles and antiparticles in the fermionic formulation.

One motivation for this study was to understand how the $SU(2)$
symmetry of the two flavor Schwinger model appears in the massless
boson field of the solution.  But a single massless boson is also
equivalent to a single free massless Dirac field.  So the latter
formulation must also have a hidden $SU(2)$ symmetry.  This is most
easily understood as a symmetry between particles and antiparticles.
For every particle state of momentum $p$ there is also the possibility
of having an antiparticle of the same momentum.  The hidden symmetry
puts these two states into a doublet.  By combining fermions with
antifermions, the $SU(2)$ symmetry does not commute with fermion
number.  A single free Dirac fermion is equivalent to an isodoublet of
Majorana fermions.

The three flavor Schwinger model is solved via its equivalence to one
massive and two massless bosons.  Thus the free theory of two massless
scalar bosons in two space-time dimensions must have a hidden $SU(3)$
symmetry.  Actually the above discussion in terms of fermions and
antifermions suggests that this is in fact a subgroup of an even
larger hidden $SU(4)$ symmetry.

\section*{Acknowledgements}
This manuscript has been authored under contract number
DE-AC02-98CH10886 with the U.S.~Department of Energy.  Accordingly,
the U.S. Government retains a non-exclusive, royalty-free license to
publish or reproduce the published form of this contribution, or allow
others to do so, for U.S.~Government purposes.  I thank Mark Rudner
for pointing out the connection between the Gompertz constant and $C$
appearing in Eq.~(\ref{propagator}).

\end{document}